\documentclass[JHEP,floatfix,amsmath,nofootinbib,superscriptaddress,amssymb,preprintnumbers,floatfix]{revtex4-1}
 \usepackage{amsmath,txfonts,longtable,booktabs,overpic,amssymb,bm,bbm,multirow,float,graphicx,color,dcolumn,subfigure,hyperref,orcidlink,tikz,fancyhdr}
\definecolor{blue}{RGB}{45,48,146}
  \hypersetup{colorlinks,citecolor=blue,anchorcolor=red,menucolor=red, linkcolor=red,filecolor=red,runcolor=red,urlcolor=blue,frenchlinks=true}

\begin{document}

\title{Generalized quantum two level model and its application in astrophysics}

\author{Orkash Amat\,\orcidlink{0000-0003-3230-7587}}
\email{Correspondence Author. email: orkashamat@nju.edu.cn}
\affiliation{ School of Astronomy and Space Science, \href{https://ror.org/01rxvg760}{Nanjing University}, Nanjing 210023, China}

\author{Nurimangul Nurmamat\,\orcidlink{0000-0001-9227-3716}}
\affiliation{ School of Astronomy and Space Science, \href{https://ror.org/01rxvg760}{Nanjing University}, Nanjing 210023, China}

\author{Yong-Feng Huang \,\orcidlink{0000-0001-7199-2906}}
\email{Correspondence Author. email: hyf@nju.edu.cn}
\affiliation{ School of Astronomy and Space Science, \href{https://ror.org/01rxvg760}{Nanjing University}, Nanjing 210023, China}
\affiliation{ Key Laboratory of Modern Astronomy and Astrophysics (\href{https://ror.org/01rxvg760}{Nanjing University}), Ministry of Education, China}

\author{Cheng-Ming Li\,\orcidlink{0000-0002-9159-8129}}
\affiliation{ School of Physics and Microelectronics, \href{https://ror.org/04ypx8c21}{Zhengzhou University}, Zhengzhou 450001, China}

\author{Jin-Jun Geng\,\orcidlink{0000-0001-9648-7295}}
\affiliation{ \href{https://ror.org/02eb4t121}{Purple Mountain Observatory}, \href{https://ror.org/034t30j35}{Chinese Academy of Sciences}, Nanjing 210023, China}

\author{Chen-Ran Hu\,\orcidlink{0000-0002-5238-8997}}
\affiliation{ School of Astronomy and Space Science, \href{https://ror.org/01rxvg760}{Nanjing University}, Nanjing 210023, China}

\author{Ze-Cheng Zou\,\orcidlink{0000-0002-6189-8307}}
\affiliation{ School of Astronomy and Space Science, \href{https://ror.org/01rxvg760}{Nanjing University}, Nanjing 210023, China}

\author{Xiao-Fei Dong\,\orcidlink{0009-0000-0467-0050}}
\affiliation{ School of Astronomy and Space Science, \href{https://ror.org/01rxvg760}{Nanjing University}, Nanjing 210023, China}

\author{Chen Deng\,\orcidlink{0000-0002-2191-7286}}
\affiliation{ School of Astronomy and Space Science, \href{https://ror.org/01rxvg760}{Nanjing University}, Nanjing 210023, China}

\author{Fan Xu\,\orcidlink{0000-0001-7943-4685}}
\affiliation{ Department of Physics, \href{https://ror.org/05fsfvw79}{Anhui Normal University}, Wuhu, Anhui 241002, China}

\author{Xiao-li Zhang\,\orcidlink{0000-0002-3877-9289}}
\affiliation{ Department of Physics, \href{https://ror.org/01rxvg760}{Nanjing University}, Nanjing 210093, China}

\author{Chen Du\,\orcidlink{0009-0002-8460-1649}}
\affiliation{ School of Astronomy and Space Science, \href{https://ror.org/01rxvg760}{Nanjing University}, Nanjing 210023, China}

\begin{abstract}
Complicated time-dependent curved spacetime and electric field are
involved in many astrophysical situations, including the early
universe, Hawking radiation, the Schwinger effect, and
gravitational pair production. In this Letter, a generalized
quantum two-level model (GQTLM) is developed, which is applicable
to arbitrary time-dependent curved spacetime and electric field.
The model is found to be consistent with quantum kinetic
theory, and is characterized by its simplicity and versatility.
The momentum distribution of particles and the effects of
gravitational distortions can be correctly described. Quantum
properties concerning vortex structures, such as the intrinsic
orbital angular momentum of particles and antiparticles can also
be conveniently calculated. The model is expected to significantly
advance the quantum exploration of the universe. It could refine
the prediction of primordial gravitational waves and relevant
non-Gaussian signals, extend the calculation of Hawking radiation
to general black hole configurations, help to distinguish neutron
stars from strange quark stars, and elucidate the gravitational
pair production mechanism.
\end{abstract}
\maketitle
\section{Introduction}
Quantum two-level model (QTLM)
\cite{Piazza:2001,Avetissian:2002ucr,Akkermans:2011yn,Kaminski:2018ywj,Krajewska:2019vqd,Fiordilino:2021zkp,Dunne:2022zlx,Bechler:2023kjx}
is a useful tool that can describe the ionization of atoms and
molecules \cite{Keldysh:1965ojf,2005PhRvL,Krausz:2009zz},
time-dependent tunneling \cite{Keski-Vakkuri:1996lbi},
Landau-Zener effect
\cite{zueco2008landau,oka2009nonequilibrium,Shevchenko:2010ms},
driven atomic system \cite{li2010carrier,jha2011experimental},
chemical reactions
\cite{miller1968semiclassical,saha2011tunneling}, Hawking
radiation \cite{Brout:1995rd,Parikh:1999mf}, cosmological particle
production \cite{Parker:1968mv}, heavy ion collisions
\cite{Greiner1985-qt,Blaschke:2017igl}, shot noise in tunnel
junctions \cite{Klich:2008un,Goren}, dynamical Casimir effect
\cite{Jaekel:1997hr,Dodonov:2010zza}, Schwinger
effect~\cite{Krajewska:2019vqd,Majczak:2024hmt}, superconductor
material~\cite{Solinas:2020woq}, gravitational
atom~\cite{Boskovic:2024fga,Tomaselli:2024faa} and so on. However,
the application of QTLM in astrophysics and particle physics is
still troubled by several limitations and drawbacks. For instance,
very complicated curved spacetime and electric field are involved
in various astrophysics situations, such as the early
universe~\cite{Huang:1970iq}, Hawking
radiation~\cite{Hawking:1974rv,Hawking:1975vcx,Gibbons:1976ue,Dumlu:2020wvd,Trevisan:2024jvn},
Schwinger
effect~\cite{Schwinger:1951nm,Bialynicki-Birula:1991jwl,Alkofer:2001ik,Hebenstreit:2011wk,Amat:2022uxq,Kohlfurst:2021skr,Ilderton:2021zej,Amat:2023vwv,Aleksandrov:2024rsz},
and gravitational pair
production~\cite{Litim:2007iu,Wondrak:2023zdi}. But the QTLM is
usually discussed in context of much simplified spacetime
conditions. Similarly, in particle physics, the QTLM is only
applicable for studying quantum phenomena in linearly polarized
electric fields and is inapplicable for situations involving
arbitrary electric fields. The generalized quantum two-level
model (GQTLM) in complicated arbitrary curved spacetime and
electric field remains unresolved. Although the
Wentzel-Kramers-Brillouin (WKB)-like QTLM has been proposed for a
supplement
~\cite{Hebenstreit:2009km,Dumlu:2010ua,Enomoto:2022mti}, the exact
QTLM for arbitrary electric fields remains unsettled~\cite{Amat:2024nvg,Amat:2025zep}.

In this paper, we propose a novel exact GQTLM to study complicated astrophysical situations for the first time.
The model is then engaged to investigate the early universe,
Hawking radiation, the Schwinger effect, and gravitational pair
production. Our model aims to address key limitations and
shortcomings of the current QTLM, ensuring its effectiveness
in arbitrary time-dependent curved spacetime and electric field.
Notably, our model incorporates spacetime distortion effects,
since the gravitational metric tensor in our model is not limited
to specific forms, offering unprecedented flexibility in
describing relevant phenomena. Secondly, the connection between
our exact GQTLM and quantum kinetic theory (QKT) is
explored, which clarifies the quantum nature of the particles
created in time-dependent curved spacetime and electric field.
Thirdly, the quantum vortex structure of the particles and
antiparticles is also studied, shedding light on their intricate
dynamics and unique properties. Since the GQTLM encompasses a
large amount of quantum information, including single particle
momentum distribution~\cite{Fedotov:2022ely}, vortex structures
(such as the particle spin angular momentum, intrinsic orbital
angular momentum, extrinsic orbital angular momentum, spin-orbit
interaction,
ect)~\cite{Allen:1992zz,Mair:2001fyp,Jentschura:2010ap,PhysRevLett.107.174802,Bahrdt:2013eoa,Bliokh:2015yhi,RevModPhys.89.035004,Bliokh:2017uvr,Ivanov:2019vxe,Ivanov:2022jzh,Karlovets:2022evc,Lu:2023wrf,Ababekri:2022mob,Ababekri:2024cyd,Ababekri:2024glc,Session:2023hoq},
tunnelling and multiphoton absorption
effects~\cite{Amat:2023vwv,Amat:2024nvg}, and effective mass
signatures~\cite{Kohlfurst:2013ura}, it has a wide application for
quantum phenomena in both astronomy and particle physics.
Additionally, our exact GQTLM may provide valuable insights into
the vacuum structure under curved spacetime and electric field.

\begin{figure*}[!t]
\setlength{\abovecaptionskip}{0.cm}
\setlength{\belowcaptionskip}{-0.cm}
\centering\includegraphics[width=1\linewidth]{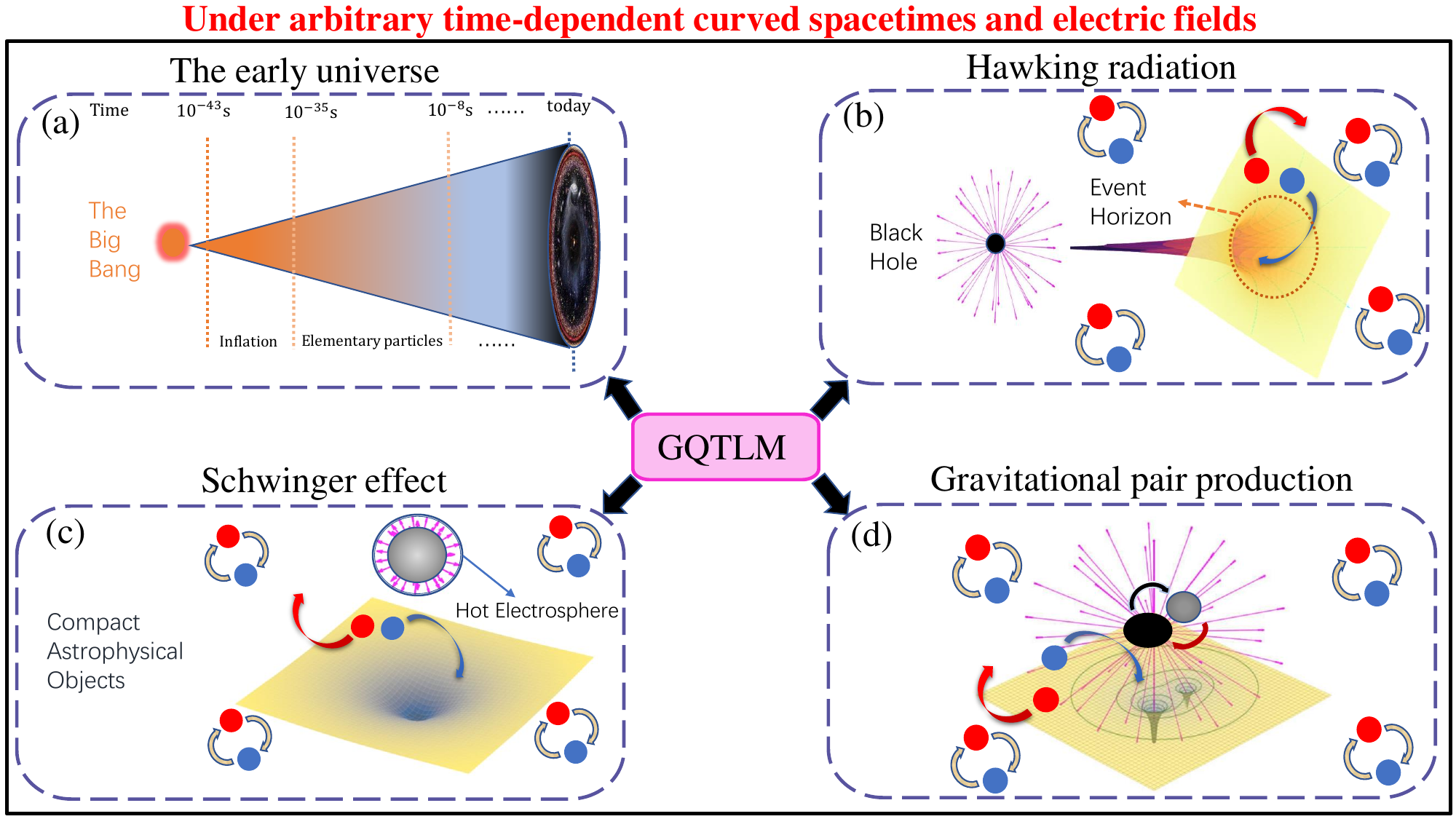}
\vspace{-0.6 cm}
\begin{picture}(300,25)
\end{picture}
\caption{Various situations involving complicated arbitrary
time-dependent curved spacetimes and electric fields: (a) the
early universe; (b) Hawking radiation; (c) the Schwinger effect;
and (d) gravitational pair production.  The red and blue balls in
each panel represent the particle and antiparticle coming from the
vacuum, respectively. The magenta lines in Panels (b), (c) and (d)
represent electric field. (a) The universe expanded exponentially quickly during the period between $10^{-43}\,{\rm s}$ and $10^{-35}\,{\rm
s}$ after the Big Bang~\cite{Kolb:1981hk,Guth:2005zr}, and the inflation field is thought to have terminated around $10^{-35}\,{\rm s}$~\cite{Podolsky:2005bw}. Subsequently, the elementary particles, created from the vacuum, were accelerated in the subsequent period between $10^{-35}\,{\rm s}$ and $10^{-8}\,{\rm s}$~\cite{Tryon:1973xi,Shakeri:2019mnt}. Throughout the overall period from $10^{-43}\,{\rm s}$ and $10^{-8}\,{\rm s}$, the Schwinger effect, driven by strong curved spacetime and intense electric field, is expected to have played a dominant role~\cite{Tryon:1973xi}. (b) Under the influence
of a black hole's curved spacetime and intense electric field,
antiparticles are pulled into the black hole, while particles
escape from it. This black hole may be an unbounded and charged
Reissner-Nordstr\"om black hole~\cite{PoncedeLeon:2017usu}. In
regions near the event horizon or close to the core, the coupling
between spacetime curvature and nonlinear correction terms may
cause the electric field strength to exhibit local maxima, forming
the so-called ``ultrastrong electric field''
zones~\cite{Balart:2014cga}. (c) Intense electric fields and the
curved spacetime around a massive compact astrophysical object
(gray ball) may give rise to the Schwinger effect in its vicinity
and hot electrosphere ~\cite{Prakapenia:2023fvx}. (d) In the
merger process of a massive compact astrophysical object (either a
neutron star or a strange quark star~\cite{Geng:2015uja,Zhang:2024uco},
represented by the gray ball) with a black hole (black ball), the
varying curvature of spacetime caused by the strong gravity of the
two objects~\cite{Zou:2022wtp}, along with the strong electric
field of the black hole, may lead to the generation of
particle-antiparticle pairs. All the four situations can be
described by GQTLM.} \label{fig1}
\end{figure*}
\section{Various complicated astrophysical situations}
Fig.~\ref{fig1} illustrates four typical situations that involve
complicated arbitrary time-dependent curved spacetime and electric
field, i.e. the early universe, Hawking radiation, the Schwinger
effect, and gravitational pair production.

The first situation, shown in Fig. \ref{fig1}(a), corresponds to
the early universe, i.e. $10^{-43}{\rm s}$ -- $10^{-35}{\rm s}$
after the Big Bang. In this period, both electric and magnetic
fields are present, but the electric field is significantly
stronger than the magnetic field \cite{Gorbar:2019fpj}. Therefore,
we mainly need to focus on the effects of the electric field.
Additionally, spacetime distortion effect in this period may be
extremely significant, and the
Friedmann-Lema\^itre-Robertson-Walker (FLRW)
metric~\cite{Friedmann:1924bb,Robertson:1929zz,Lemaitre:1931zza,Robertson:1933zz,Walker:1937qxv,Weinberg:1988cp,Pettinari:2016,Kitamoto:2020tjm,Okano:2021pva,Meimanat:2023hjq,Lysenko:2023wrs}
may fail to describe it effectively. The spacetime distortion
effect is closely related to the off-diagonal elements of the
metric tensor of the gravity. To address this issue, we need to
incorporate arbitrary time-dependent curved spacetime and electric
field so as to provide a more accurate representation of the
particle creation process during this phase.

The second situation, illustrated in Fig. \ref{fig1}(b),
corresponds to the process of Hawking radiation. In traditional
Hawking radiation process, the emitted particles or antiparticles
are generally believed to originate from quantum fluctuations
caused by the black hole's gravity near the event
horizon~\cite{Hawking:1975vcx} or through tunnelling process
caused by gravity~\cite{Hawking:1975vcx,Parikh:1999mf}. This
mechanism results in an almost symmetrical number of particles and
antiparticles, thus exhibits a high degree of overall symmetry.
However, in the cases considered here, the radiated particles and
antiparticles can further come from processes driven by a strong
electric field, including both multiphoton-dominated and
tunnelling-dominated mechanisms~\cite{Amat:2024nvg}. Under the
influence of this strong electric field, the vacuum outside the
black hole's event horizon can create particle-antiparticle pairs,
which further alters the emission process. Moreover, the presence
of the strong electric field may break the symmetry between the
numbers of particles and antiparticles, leading to asymmetric
radiation features. Exploring these phenomena will help deepen our
understanding of the Hawking radiation mechanism and reveal the
profound influence of strong electric field on the quantum
dynamics near black holes.

The third situation, shown in ~Fig.~\ref{fig1}(c), corresponds to
the Schwinger effect caused by compact astrophysical objects (not
including black holes) with strong electric fields.
Particle-antiparticle pairs could be produced in the process,
leading to a strong particle-antiparticle pair winds from the hot
electrosphere. However, previous studies on the generation of the
particle-antiparticle pair winds originating from compact
astrophysical objects generally omitted the joint effect of curved
spacetime and electric field, which are in fact critical in
understanding the dynamics of the
wind~\cite{Prakapenia:2023fvx,Zou:2022voz,Prakapenia:2023tsw}. In this study,
we will consider the joint effect of curved
spacetime~\cite{Collas:2018jfx,Bar:2009zzb} and strong electric
field. A comprehensive description of the Schwinger effect will be
presented.

The fourth situation, shown in  Fig.~\ref{fig1} (d), corresponds
to the gravitational wave pair production that occurs during the
merger of a supermassive compact object with a charged black hole.
It involves the interaction between gravitational
waves~\cite{Saulson:2010zz,LIGOScientific:2016aoc,Maggiore:2008}
and electric field in an extreme astrophysical environment. During
the merger process, the intense gravitational waves and the strong
electric field could lead to the creation of particle-antiparticle
pairs from the vacuum, which reflects the combined effects of
gravity and electric field.
\section{Generalized quantum two-level model}
A common feature of
these four situations is the particle-antiparticle pair creation
from the vacuum, which is the core process underlying each
scenario. This common feature leads to a unified perspective for
understanding the phenomena. Here we develop the GQTLM to present
a unified description for the scalar vacuum pair production
process. The quantum properties of scalar particles and fermions
are nearly identical despite their difference in the distribution
function~\cite{Amat:2024nvg}. This allows us to focus on the
essential physics without introducing additional complexity. So,
we can simply focus on the scalar vacuum pair production process.

In order to deal with the complicated arbitrary time-dependent
curved spacetimes and/or electric fields involved in the early
universe, Hawking radiation, the Schwinger effect, and
gravitational pair production, we express the Lagrangian density
in the natural units ($\hbar=c=1$) as
\begin{widetext}
\begin{align}\label{eq:1}
\begin{split}
\mathcal{L}&=\begin{cases}
\sqrt{-g}\left(-\frac{M_{p}^{2}}{2}R_{1}+ \frac{1}{2}g^{\mu\nu}\partial_{\mu}\phi\partial_{\nu}\phi -V(\phi) +g^{\mu\nu}(\mathcal{D}_{\mu}\chi)^{\dagger}(\mathcal{D}_{\nu}\chi)-(m^{2}-\xi_{1} R_{1})\chi^{\dagger}\chi -\frac{1}{4}I(\phi)F_{\mu\nu}F^{\mu\nu} \right), ~~{\rm the~early~universe} \\
\sqrt{-g}\left(g^{\mu\nu}(\mathcal{D}_{\mu}\chi)^{\dagger}(\mathcal{D}_{\nu}\chi)-(m^{2}-\xi_{2} R_{2})\chi^{\dagger}\chi -\frac{1}{4}F_{\mu\nu}F^{\mu\nu} \right), ~~~~~~~~~~~~~~~~~~~~~~~~~~~~~~~~~~~~~~~~~~~~~~~~~~~~~~~~~~{\rm the~Hawking~radiation} \\
\sqrt{-g}\left( g^{\mu\nu}(\mathcal{D}_{\mu}\chi)^{\dagger}(\mathcal{D}_{\nu}\chi)-(m^{2}-\xi_{3} R_{3})\chi^{\dagger}\chi -\frac{1}{4}F_{\mu\nu}F^{\mu\nu} \right), ~~~~~~~~~~~~~~~~~~~~~~~~~~~~~~~~~~~~~~~~~~~~~~~~~~~~~~~~~~{\rm the~Schwinger~effect}\\
\sqrt{-g}\left( g^{\mu\nu}(\mathcal{D}_{\mu}\chi)^{\dagger}(\mathcal{D}_{\nu}\chi)-(m^{2}-\xi_{4} R_{4})\chi^{\dagger}\chi -\frac{1}{4}F_{\mu\nu}F^{\mu\nu} \right), ~~~~~~~~~~~~~~~~~~~~~~~~~~~~~~~~~~~~~~~~~~~~~~~~~~~~~~~~~~{\rm the~gravitational~pair~production}
\end{cases}
\end{split}
\end{align}
\end{widetext}
where $d^2 s =g_{\mu \nu}(t)dx^{\mu}dx^{\nu}$ is the metric that
can be time-dependent and $A^{\mu}=\left(0,~\mathbf{A}(t)
\right)=\left(0,~A_{x}(t),~A_{y}(t),~A_{z}(t) \right)$ is the
four-vector potential, with ${\mu,\nu}=\{0, 1, 2, 3\}$ and
$g_{\mu\nu}$ being a time-dependent matric. Here we further define
$g={\rm det}(g_{\mu\nu})$. $M_{p}=(8\pi G)^{-1/2}=2.43\times
10^{18}\,{\rm GeV}$ is the reduced Planck mass. For simplicity, we
use the index $a=\{1, 2, 3, 4\}$ to represent the situations of
the early universe, Hawking radiation, the Schwinger effect, and
gravitational pair production. The coupling constant
$\xi_{a}$ quantifies the nonminimal coupling of the scalar field
$\chi$ to gravity. In Equation~\ref{eq:1}, $R_{a}$ is the scalar
curvature in the Einstein field equation~\cite{Einstein:1916vd},
$\phi$ is a single scalar inflaton field, $V(\phi)$ denotes its
potential, $F_{\mu\nu}=\partial_\mu A_\nu-\partial_\nu A_\mu$
denotes the strength of the external fields, $I(\phi)$ is the
kinetic coupling function between the inflaton fields and $U(1)$
gauge fields, $\mathcal{D}_{\mu}=\partial_{\mu}-iq A_{\mu}$ is the
covariant derivative acting on the scalar field  $\chi$ with
charge $q$ and mass $m$. The matric takes a general form of
\begin{align}\label{eq:2}
g_{\mu \nu} =g_{\mu \nu}(t) =
\begin{pmatrix}
g_{00}(t) & g_{01}(t) & g_{02}(t) & g_{03}(t) \\
g_{10}(t) & g_{11}(t) & g_{12}(t) & g_{13}(t) \\
g_{20}(t) & g_{21}(t) & g_{22}(t) & g_{23}(t) \\
g_{30}(t) & g_{31}(t) & g_{32}(t) & g_{33}(t)
\end{pmatrix}.
\end{align}

We consider a quantum two-state system of the scalar field, described by 
$\left\lvert ~\chi \right\rangle = c_{a,~\bf{k}}^{(1)}(t) \left\lvert 1 \right\rangle + c_{a,~\bf{k}}^{(2)}(t) \left\lvert 2 \right\rangle$ in the momentum space $\bf{k}$.
Here, $\left\lvert 1 \right\rangle$ and $\left\lvert 2 \right\rangle$ are the ground and excited states of the scalar field, respectively. The coefficients $c_{a,~\bf{k}}^{(1)}(t)$ and $c_{a,~\bf{k}}^{(2)}(t)$ represent the time-dependent probability amplitudes. Using this representation, the exact GQTLM  can be derived as [see Appendices \ref{sec:a}-\ref{sec:d}]
\begin{equation}\label{eq:3}
i\frac{d}{dt}\begin{bmatrix}
c_{a,~\bf{k}}^{(1)}(t)\\
c_{a,~\bf{k}}^{(2)}(t)
\end{bmatrix}
=
\begin{pmatrix} \Omega_{a,~\bf{k}}(t) & -iQ_{a,~\bf{k}}(t) \cr -iQ_{a,~\bf{k}}(t) & -\Omega_{a,~\bf{k}}(t) \end{pmatrix}
\begin{bmatrix}
c_{a,~\bf{k}}^{(1)}(t)\\
c_{a,~\bf{k}}^{(2)}(t)
\end{bmatrix},
\end{equation}
where
\begin{align}\label{eq:4}
Q_{a,~\bf{k}}(t) &=-\frac{\dot{\Omega}_{a,~\mathbf{k}}(t)}{2\Omega_{a,~\mathbf{k}}(t)},\\ \label{eq:5}
\Omega_{a,~\mathbf{k}}(t) &=\sqrt{\mathcal{E}^2_{a}(t) - \frac{1}{2} \dot{P}(t) - \frac{1}{4} P^2(t)},\\ \label{eq:6}
\mathcal{E}^2_{a}(t) &=
\left(
\frac{3\dot{g}^2-4\ddot{g}}{16g^2}
- \frac{\dot{g}}{4g} \cdot \frac{\dot{g}^{00}}{g^{00}}
+ \frac{\dot{g}^{0j}}{g^{00}}(-i k_j - i e A_j)
\right)
\nonumber \\
&\quad + \frac{\dot{g}^{0j}}{g^{00}}(-i e \partial_0 A_j) + \frac{g^{i0}}{g^{00}}
\left(
\frac{\dot{g}}{4g} i k_i + i e A_i \cdot \frac{\dot{g}}{4g}
\right)
\nonumber \\
&\quad + g^{ij}
\left(
-k_j k_i - e A_j k_i - e A_i k_j - e^2 A_i A_j
\right) \nonumber \\
&\quad + \frac{g^{0j}}{g^{00}}
\left(
\frac{\dot{g}}{2g} (-i k_j)
\right) + \left(
m^2 - \xi_{a} R_{a}
\right),
\end{align}
\begin{align} \label{eq:7}
P(t) &= \frac{\dot{g}^{00}}{g^{00}}
+ \frac{g^{0j}}{g^{00}}(-i k_j)
+ \frac{g^{0j}}{g^{00}}(-i e A_j)
+ \frac{g^{i0}}{g^{00}}(-i k_i) \nonumber \\
&\quad + \frac{g^{i0}}{g^{00}}(-i e A_i).
\end{align}
Here $\Omega_{a,~\mathbf{k}}(t)$ denotes the single particle energy in
the curved spacetime and electric field. The initial condition is
featured by $c_{a,~\bf{k}}^{(1)}(t_0)=1$ and
$c_{a,~\bf{k}}^{(2)}(t_0)=0$, omitting the effects of the curved
spacetime and electric field. The scalar particle momentum
distribution can be calculated by using the coefficient
$c_{a,~\bf{k}}^{(2)}(t)$ as
\begin{eqnarray}\label{eq:8}
f_{a,~\bf k}(t)=2|c_{a,~\bf{k}}^{(2)}(t)|^2.
\end{eqnarray}

For the diagonalized FLRW metric, our GQTLM reduces to the WKB
result of Ref.~\cite{Sobol:2020frh}. When only the effect of the
electric field is included, it reduces to normal two level
model~\cite{Amat:2024nvg}. The reliability and accuracy of the
GQTLM is thus further demonstrated.
\section{Quantum kinetic theory}
The quantum Vlasov equation
(QVE) and the Wigner function are collectively called QKT. Note
that the GQTLM is also consistent with QKT, as detailed below. The
QVE of scalar vacuum pair production for the early universe,
Hawking radiation, Schwinger effect and gravitational pair
production in a curved spacetime and electric field can be
obtained as [see the Appendix \ref{sec:e}]
\begin{eqnarray}
\dot{\mathcal{F}}_{a,~\mathbf{k}}&=&\frac{1}{2} W_{a,~\bf{k}}(t)~\mathcal{G}_{a,~\mathbf{k}},\label{eq:9}\\
\dot{\mathcal{G}}_{a,~\mathbf{k}}&=&W_{a,~\bf{k}}(t)~(1+2\mathcal{F}_{a,~\mathbf{k}})-2\Omega_{a,~\mathbf{k}}\mathcal{H}_{a,~\mathbf{k}},\label{eq:10}\\
\dot{\mathcal{H}}_{a,~\mathbf{k}}&=&2\Omega_{a,~\mathbf{k}}~\mathcal{G}_{a,~\mathbf{k}}.\label{eq:11}
\end{eqnarray}
where
\begin{align}\label{eq:12}
W_{a,~\bf{k}}(t)=\frac{1}{2} Q_{a,~\bf{k}}(t).
\end{align}
Omitting the effects of curved spacetime and electric field, the
initial condition can be expressed as
\begin{equation}\label{eq:13}
\mathcal{F}_{a,~\mathbf{k}}(t_0)=\mathcal{G}_{a,~\mathbf{k}}(t_0)=\mathcal{H}_{a,~\mathbf{k}}(t_0)=0.
\end{equation}

The scalar particle momentum distribution is
\begin{eqnarray}\label{eq:14}
f_{a,~\bf k}(t)=\mathcal{F}_{a,~\mathbf{k}}(t).
\end{eqnarray}
The number density of created particles can be calculated by
integrating the momentum distribution function
$\mathcal{F}_{a,~\mathbf{k}}(t)$ with respect to the momentum
$\mathbf{k}$:
\begin{equation}\label{eq:15}
n_{a,~\mathbf{k}}(t)=\int \frac{d^3k}{(2\pi)^3}f_{a,~\mathbf{k}}(t).
\end{equation}
It demonstrates that our GQTLM is equivalent to QVE. For the
diagonalized FLRW metric, the equation simplifies to the results
shown in Eqs.~(19), (20), and (21) of Ref.~\cite{Sobol:2020frh}.
Furthermore, when only consider the effect of the electric field,
our model also aligns with previous relevant
studies~\cite{Amat:2024nvg}.

The advantage of QVE is that it can explain chirping, tunneling,
multiphoton, and interference effects~\cite{Hebenstreit:2009km}.
The equivalence between our GQTLM and QVE means that our model is
also capable of explaining these effects in the presence of
spacetime curvature and electric field.
\section{Origin of intrinsic orbital angular momentum}
Our GQTLM
can not only describe the momentum distribution of particles but
also provide extensive quantum information for the particle
creation process under the influence of spacetime curvature and
electric field. For instance, the topological charge or the
intrinsic orbital angular momentum quantum number $l$ of the
created particles and antiparticles can be determined~\cite{Amat:2025zep}. In fact,
$l$ can be calculated as
\begin{align}\label{eq:16}
l=\frac{\phi_{\rm Berry}}{2\pi}=\frac{1}{2\pi}\int_\mathcal{C} {\mathcal{A}}\cdot d\bf{k},
\end{align}
where $\phi_{\rm Berry}$ is the Berry phase~\cite{Berry:1984jv}.
In the momentum space ($\bf{k}$), the gauge-invariant Berry
connection of the complex function ($c_{a,~\bf{k}}^{(2)}(t)$) is
\begin{align}\label{eq:17}
{\mathcal{A}} = \frac{\text{Re} \left(c_{a,~\bf{k}}^{*(2)}(t)\left[-i\nabla_{\boldsymbol{k}}\right]c_{a,~\bf{k}}^{(2)}(t)\right)}{\lvert c_{a,~\bf{k}}^{(2)}(t)\rvert^2} = \nabla_{\boldsymbol{k}}\left(\arg \left[c_{a,~\bf{k}}^{(2)}(t)\right]\right),
\end{align}
where the term ``arg'' represents the argument of the complex
function $c_{a,~\bf{k}}^{(2)}(t)$, i.e., $\arg
\left[c_{a,~\bf{k}}^{(2)}(t)\right]= {\rm arctan}\left(
\Im\left[c_{a,~\bf{k}}^{(2)}(t)\right]/\Re\left[c_{a,~\bf{k}}^{(2)}(t)\right]
\right)$. It indicates that the intrinsic orbital angular momentum
of particles can be interpreted through the GQTLM. In short, the
GQTLM is an informative tool. It encompasses the momentum
distribution, intrinsic orbital angular momentum, and other
related properties of particles created in curved spacetime and
electric field.
\section{Outlook}
While the cosmic inflation theory excels in
explaining the anisotropy and spectral index of the cosmic
microwave background (CMB), some uncertainties remain in its
specific predictions~\cite{Penzias:1965wn}. For example, the
amplitude of primordial gravitational waves is yet to be
determined, and the predicted non-Gaussian signals still could not
be detected~\cite{BICEP2:2014owc}. Addressing these issues is
crucial for refining the inflation theory and validating its
specific predictions (Fig.~\ref{fig1}a). Our GQTLM model provides
strong support for the inflation as well as a fresh perspective
for understanding the process. It lays a theoretical foundation
for uncovering the essence of the inflation and points the way for
future in-depth research.

Hawking radiation is interpreted by adopting quantum field theory
in curved spacetime, but its microscopic physical mechanism
remains unclear~\cite{Parikh:1999mf}. For example, the exact
process of particle generation and the underlying dynamics are not
fully understood. Additionally, experimental or observational test
of the Hawking radiation mechanism is still a challenge. For
higher-dimensional, charged or rotating black holes, the
computation and analysis of Hawking radiation are more complex.
Extending the theory to encompass more general black hole models
is an important area of ongoing research~\cite{Ida:2002ez}. Our
GQTLM could help address these issues and may have the potential
to extend the study of Hawking radiation to broader black hole
configurations (Fig.~\ref{fig1}b). It provides a fresh perspective
on uncovering the mysteries of black hole physics and opens up new
possibilities for related researches.

Distinguishing neutron stars from strange quark stars remains a
longstanding challenge in astrophysics from both theoretical and
observational perspectives~\cite{Zhang:2024xod}. Attempts have been made based on their
difference in the mass-radius relationship, cooling
speed~\cite{Huang:1997,Madsen:1998qb}, limiting rotation
period~\cite{Glendenning:1989cf}, high energy
bursts~\cite{Dai:2015jwa}, and gravitational wave radiation
~\cite{Madsen:1998qb}. However, none of these methods is effective
enough. Developing more effective methods is still a challenging
task. Note that the electron-positron pair winds emanating from
these two kinds of objects may be very different
(Fig.~\ref{fig1}c), which can be calculated by using our GQTLM.
The electric field generated by the hot electrosphere of a strange
quark star is much stronger than that of a neutron star. As a
result, the electron-positron pair winds of strange quark stars
are more intense than that of neutron stars. This characteristics
could help us distinguish the two types of compact stars.

Theoretically, strong gravitational waves can polarize and
interact with quantum fields in a vacuum, leading to the
generation of particle-antiparticle pairs~\cite{Ford:1986sy}.
However, the detailed dynamics of this process, the response of
the quantum fields, and the coherence conditions are not yet fully
understood (~Fig.~\ref{fig1}d). Our GQTLM could offer profound
theoretical insights into these challenging problems. It can help
uncover the fundamental mechanisms behind the issue.

The GQTLM may also be applied in other fields, such as quantum
Hall effect. It can explain particle vortex phenomena, which
further enhances its theoretical relevance~\cite{Wu:2019kyb}.

Despite its advantages, the GQTLM has several limitations that
warrant further theoretical exploration. First, it does not account for the effects of the magnetic field and spatiotemporal dependence of the gravitational metric tensor, represented as $g_{\mu \nu}=g_{\mu \nu}(t,x)$.
Second, some effects predicted by the model are extremely
difficult to test through astronomical observations, posing a
clear challenge for experimental verification.
\section{Conclusions}
In this study, a generalized quantum two
level model is developed to describe the pair creation process
involving complicated arbitrary time-dependent curved spacetime
and electric field. It can be applied to various astrophysical
situations such as the early universe, Hawking radiation, the
Schwinger effect, and gravitational pair production. The model is
shown to be consistent with QKT, highlighting its effectiveness in
describing these phenomena. The GQTLM exhibits significant
advantages in the following aspects. First, it accurately
characterizes the momentum distribution of particles created in
arbitrary time-dependent curved spacetime and electric fields.
Second, it effectively incorporates the impact of gravitational
field on particle dynamics, such as the gravitational distortion
effect. Third, the GQTLM model conveniently describes the
particles' quantum properties involving vortex structures, such as
the intrinsic orbital angular momentum by engaging the Berry
phase.
\section{Acknowledgments}
This work was supported by the National
Natural Science Foundation of China (NSFC) under Grant Nos. 12447179, 12233002, 12005192, 12273113.  YFH was also supported by National
SKA Program of China No. 2020SKA0120300, by the National Key R\&D
Program of China (2021YFA0718500), and by the Xinjiang Tianchi
Program. Cheng-Ming Li was also supported by the Natural Science Foundation of Henan Province of China (No. 242300421375) and by the Project funded by China Postdoctoral Science Foundation (Grant Nos. 2020TQ0287 and 2020M672255). Jin-Jun Geng acknowledges support from the Youth Innovation Promotion Association (2023331).

\appendix

\section{Generalized quantum two-level model in context of the early universe}
\label{sec:a}

The exact generalized quantum two-level model (GQTLM) can be used
to study the scalar vacuum pair production in the early universe.
To deal with the complicated time-dependent curved spacetime and
electric field, we start from the Lagrangian density in the
Einstein representation~\cite{Sobol:2020frh}
\begin{align}\label{eq:s1}
\mathcal{L}_{1}=\sqrt{-g}\left[-\frac{M_{p}^{2}}{2}R_{1}+ \mathcal{L}^{\rm scalar}_{\rm inflaton} +\mathcal{L}^{\rm scalar}_{\rm QED}\right],
\end{align}
where ${\mu,\nu}=\{0, 1, 2, 3\}$, $g_{\mu\nu}$ is an arbitrary
time-dependent matric, $g={\rm det}(g_{\mu\nu})$, $M_{p}=(8\pi
G)^{-1/2}=2.43\times 10^{18}\,{\rm GeV}$ is the reduced Planck
mass, and $R_{1}$ is the scalar curvature. The inflationary term
can be expressed by engaging a single scalar inflation field
$\phi$ with a potential of $V(\phi)$,
\begin{align}\label{qe:s2}
\mathcal{L}^{\rm scalar}_{\rm inflaton}&=\frac{1}{2}g^{\mu\nu}\partial_{\mu}\phi\partial_{\nu}\phi -V(\phi).
\end{align}
The Lagrangian density of the scalar quantum electrodynamics (QED) is
\begin{align}\label{eq:s3}
\mathcal{L}^{\rm scalar}_{\rm QED}& = \mathcal{L}_{\rm EM} +
\mathcal{L}^{\rm scalar}_{\rm ch},
\end{align}
where the electromagnetic and charged scalar fields can be further
expressed as
\begin{align}\label{eq:s4}
\mathcal{L}_{\rm EM}&=-\frac{1}{4}F_{\mu\nu}F^{\mu\nu} +
\mathcal{L}_{\rm
int}(A_{\mu},\phi) = -\frac{1}{4}I(\phi)F_{\mu\nu}F^{\mu\nu},\\
\label{eq:s5} \mathcal{L}^{\rm scalar}_{\rm ch}& =
g^{\mu\nu}(\mathcal{D}_{\mu}\chi)^{\dagger}(\mathcal{D}_{\nu}\chi)-(m^{2}-\xi_{1}
R_{1})\chi^{\dagger}\chi.
\end{align}
Here $F_{\mu\nu}=\partial_\mu A_\nu-\partial_\nu A_\mu$ denotes
the field strength of the external field, $\mathcal{L}_{\rm int}$
describes the interaction of the electromagnetic field with the
inflation that leads to the generation of electromagnetic field,
$I(\phi)$ is the kinetic coupling function between the inflation
field and the $U(1)$ gauge field, $\mathcal{D}_{\mu} =
\partial_{\mu} - iq A_{\mu}$ is the covariant derivative acting on
the scalar field $\chi$ with a charge $q$ and mass $m$. The
coupling constant $\xi_{1}$ quantifies the nonminimal coupling of
the scalar field $\chi$ with gravity~\cite{Sobol:2020frh}.

The dynamics of the scalar field $\chi$ can be obtained by using
the Euler-Lagrange equation
\begin{align}\label{eq:s6}
\frac{1}{\sqrt{-g}}\mathcal{D}_{\mu}\left[\sqrt{-g}g^{\mu\nu}\mathcal{D}_{\nu}\chi\right]+(m^{2}-\xi_{1} R_{1})\chi=0.
\end{align}
The solution of the scalar field $\chi$ for
$A^{\mu}=(0,\mathbf{A}(t))$ and ${\rm div\,}\mathbf{A}=0$ can be
written as
\begin{align}\label{eq:s7}
\chi(t,\mathbf{x})=\int \frac{d^{3}\mathbf{k}}{(2\pi)^{3/2}(-g)^{1/4}}\left[\hat{a}_{\mathbf{k}}\Phi_{\mathbf{k}}(t)e^{i\mathbf{k}\cdot\mathbf{x}}+\hat{b}_{\mathbf{k}}^{\dagger}\Phi^{*}_{-\mathbf{k}}(t)e^{-i\mathbf{k\cdot\mathbf{x}}}\right],
\end{align}
where the creation ($\hat{a}_{\mathbf{k}}^{\dagger}$, $\hat{b}_{\mathbf{k}}^{\dagger}$) and annihilation ($\hat{a}_{\mathbf{k}}$, $\hat{b}_{\mathbf{k}}$)  operators satisfy the standard bosonic commutation relations, i.e.,
\begin{align}\label{eq:s8}
[\hat{a}_{\mathbf{k}},\,\hat{a}^{\dagger}_{\mathbf{p}}]=[\hat{b}_{\mathbf{k}},\,\hat{b}^{\dagger}_{\mathbf{p}}]=\delta^{3}(\mathbf{k}-\mathbf{p}).
\end{align}

Adopting the transformation of
$\Phi_{\mathbf{k}}(t)=\varphi_{\mathbf{k}}(t)
e^{-\frac{1}{2}\int^t P(\tau) d\tau}$, we can obtain a
Schr\"{o}dinger like equation by substituting Eq. \eqref{eq:s7}
into Eq. \eqref{eq:s6},
\begin{align}\label{eq:s9}
\ddot{\varphi}_{\mathbf{k}}(t) +
\Omega_{\mathbf{k}}^{2}(t)\varphi_{\mathbf{k}}(t)=0,
\end{align}
where
\begin{align}\label{eq:s10}
\Omega_{\mathbf{k}}(t) &=\sqrt{\mathcal{E}^2(t) - \frac{1}{2} \dot{P}(t) - \frac{1}{4} P^2(t)},\\ \label{eq:s11}
\mathcal{P}(t) &=
\left[
  \frac{\dot{g}^{00}}{g^{00}}
+ \frac{g^{0j}}{g^{00}}(-i k_j)
+ \frac{g^{0j}}{g^{00}}(-i e A_j)
+ \frac{g^{i0}}{g^{00}}(-i k_i)
+ \frac{g^{i0}}{g^{00}}(-i e A_i)
\right], \\ \label{eq:s12}
\mathcal{E}^2(t) &=
\left[
  \frac{3\dot{g}^2-4\ddot{g}}{16g^2}
- \frac{\dot{g}}{4g} \cdot \frac{\dot{g}^{00}}{g^{00}}
+ \frac{\dot{g}^{0j}}{g^{00}}(-i k_j - i e A_j)
\right]
+ \frac{\dot{g}^{0j}}{g^{00}}(-i e \partial_0 A_j) \nonumber \\
&\quad + \frac{g^{i0}}{g^{00}}
\left(
\frac{\dot{g}}{4g} i k_i + i e A_i \cdot \frac{\dot{g}}{4g}
\right)
+ \frac{g^{0j}}{g^{00}}
\left(
\frac{\dot{g}}{2g} (-i k_j)
\right) \nonumber \\
&\quad + g^{ij}
\left(
-k_j k_i - e A_j k_i - e A_i k_j - e^2 A_i A_j
\right)
+ \left(
m^2 - \xi_{1} R_{1}
\right).
\end{align}

We now apply the Bogoliubov transformation by defining $\alpha^{'}_{\bf{k}}(t)$ and $\beta^{'}_{\bf{k}}(t)$ as
\begin{align}
\label{eq:s13}
\varphi_{\mathbf{k}}(t) =
\frac{\alpha^{'}_{\bf{k}}(t)}{\sqrt{2
\Omega_{\mathbf{k}}(t)}}e^{-i\int^t d\tau
\Omega_{\mathbf{k}}(\tau)}+\frac{\beta^{'}_{\bf{k}}(t)}{\sqrt{2
\Omega_{\mathbf{k}}(t)}}e^{i\int^t d\tau
\Omega_{\mathbf{k}}(\tau)},
\end{align}
then we have
\begin{align}
\label{eq:s14}
  \Phi_{\bf{k}}(t)& = \frac{\alpha_{\bf{k}}(t)}{\sqrt{2 \Omega_{\mathbf{k}}(t)}}e^{-i\int^t
  d\tau \Omega_{\mathbf{k}}(\tau)}
  +\frac{\beta_{\bf{k}}(t)}{\sqrt{2 \Omega_{\mathbf{k}}(t)}}e^{i\int^t d\tau \Omega_{\mathbf{k}}(\tau)},\\
  \label{eq:s15}
\dot{\Phi}_{\bf{k}}(t)& =
-i\Omega_{\mathbf{k}}(t)\left(\frac{\alpha_{\bf{k}}(t)}{\sqrt{2
\Omega_{\mathbf{k}}(t)}}e^{-i\int^t d\tau
\Omega_{\mathbf{k}}(\tau)}-\frac{\beta_{\bf{k}}(t)}{\sqrt{2
\Omega_{\mathbf{k}}(t)}}e^{i\int^t d\tau
\Omega_{\mathbf{k}}(\tau)}\right),
\end{align}
where $\alpha_{\bf{k}}(t) =
\alpha^{'}_{\bf{k}}(t)e^{-\frac{1}{2}\int^t P(\tau) d\tau}$ and
$\beta_{\bf{k}}(t) = \beta^{'}_{\bf{k}}(t)e^{-\frac{1}{2}\int^t
P(\tau) d\tau}$. It is easy to further get
\begin{eqnarray}
\label{eq:s16}
\dot{\alpha}_{\bf{k}}(t)&=&\frac{\dot{\Omega}_{\bf{k}}(t)}{2\Omega_{\mathbf{k}}(t)}\beta_{\bf{k}}(t)e^{2i\int^t d\tau \Omega_{\mathbf{k}}(\tau)},\\ \label{eq:s17}
\dot{\beta}_{\bf{k}}(t)&=&\frac{\dot{\Omega}_{\bf{k}}(t)}{2\Omega_{\mathbf{k}}(t)}\alpha_{\bf{k}}(t)e^{-2i\int^t d\tau \Omega_{\mathbf{k}}(\tau)}.
\end{eqnarray}
The Bogoliubov transformation changes the time-independent bases
of creation $a_{\bf k}$ and annihilation $b_{-{\bf k}}^\dagger$
operators to the time-dependent bases of creation $\tilde{a}_{\bf
k}(t)$ and annihilation $\tilde{b}_{-{\bf k}}^\dagger(t)$
operators, through a linear transformation of
\begin{eqnarray}\label{eq:s18}
\begin{pmatrix}
\tilde{a}_{\bf k}(t)\cr
\tilde{b}_{-{\bf k}}^\dagger(t)
\end{pmatrix}
=\begin{pmatrix}
\alpha_{\bf k} & \beta_{\bf k}^*\cr
\beta_{\bf k} & \alpha_{\bf k}^*
\end{pmatrix}
\begin{pmatrix}
a_{\bf k}\cr
b^\dagger_{-{\bf k}}
\end{pmatrix},
\end{eqnarray}
where the bosonic commutation relations are preserved by $|\alpha_{\bf{k}}(t)|^2 -|\beta_{\bf{k}}(t)|^2=1$.

Now we introduce two useful new coefficients as
\begin{eqnarray}
\label{eq:s19}
c^{(1)}_{\bf{k}}(t)&=&\beta_{\bf{k}}(t)e^{-i\int^t d\tau
\Omega_{\mathbf{k}}(\tau)},\\ \label{eq:s20}
c^{(2)}_{\bf{k}}(t)&=&\alpha_{\bf{k}}(t)e^{i\int^t d\tau
\Omega_{\mathbf{k}}(\tau)}.
\end{eqnarray}
After taking the derivatives of time on the above equations, we
can obtain
\begin{equation}\label{eq:s21}
i\frac{d}{dt}\begin{bmatrix}
c_{\bf{k}}^{(1)}(t)\\
c_{\bf{k}}^{(2)}(t)
\end{bmatrix}
=
\begin{pmatrix} \Omega_{\bf{k}}(t) & -iQ_{\bf{k}}(t) \cr -iQ_{\bf{k}}(t) & -\Omega_{\bf{k}}(t) \end{pmatrix}
\begin{bmatrix}
c_{\bf{k}}^{(1)}(t)\\
c_{\bf{k}}^{(2)}(t)
\end{bmatrix},
\end{equation}
where
\begin{align}\label{eq:s22}
Q_{\bf{k}}(t)=-\frac{\dot{\Omega}_{\mathbf{k}}(t)}{2\Omega_{\mathbf{k}}(t)}.
\end{align}
The initial conditions are $c_{\bf{k}}^{(1)}(t_0)=1$ and
$c_{\bf{k}}^{(2)}(t_0)=0$. The scalar particle momentum
distribution can be calculated by using the coefficient
$c_{\bf{k}}^{(2)}(t)$ as
 \begin{eqnarray}\label{eq:s23}
f_{\bf k}(t)=2|c_{\bf{k}}^{(2)}(t)|^2.
\end{eqnarray}

\section{GQTLM in context of the Hawking radiation}
\label{sec:b}

Hawking radiation is a quantum field phenomenon arising from
quantum fluctuations near the event horizon of a black hole, where
intense gravitational field leads to the creation of
particle-antiparticle pairs~\cite{Hawking:1974rv}. Additionally,
the strong electric field around a charged black hole may also
induce pair production, converting vacuum fluctuations into real
particle-antiparticle pairs. To explore the combination of these
two ingredients, we need the GQTLS that takes into account both
the Hawking radiation and the Schwinger effect. For this purpose,
the scalar QED Lagrangian density in arbitrary time-dependent
curved spacetime and electric field is expressed as
\begin{align}\label{eq:s24}
\mathcal{L}_{2}&=\sqrt{-g}\left[-\frac{1}{4}F_{\mu\nu}F^{\mu\nu} + g^{\mu\nu}(\mathcal{D}_{\mu}\chi)^{\dagger}(\mathcal{D}_{\nu}\chi)-(m^{2}-\xi_{2} R_{2})\chi^{\dagger}\chi \right],
\end{align}
where $g^{\mu\nu}$ is an arbitrary time-dependent matric.

Similar to the above section, the exact GQTLM in an arbitrary
time-dependent curved spacetime and electric field can be
characterized by
\begin{equation}
\label{eq:s25}
i\frac{d}{dt}\begin{bmatrix}
c_{\bf{k}}^{(1)}(t)\\
c_{\bf{k}}^{(2)}(t)
\end{bmatrix}
=
\begin{pmatrix} \Omega_{\bf{k}}(t) & -iQ_{\bf{k}}(t) \cr -iQ_{\bf{k}}(t) & -\Omega_{\bf{k}}(t) \end{pmatrix}
\begin{bmatrix}
c_{\bf{k}}^{(1)}(t)\\
c_{\bf{k}}^{(2)}(t)
\end{bmatrix},
\end{equation}
where
\begin{align}\label{eq:s26}
Q_{\bf{k}}(t)=-\frac{\dot{\Omega}_{\mathbf{k}}(t)}{2\Omega_{\mathbf{k}}(t)}.
\end{align}
The initial condition is $c_{\bf{k}}^{(1)}(t_0)=1$ and
$c_{\bf{k}}^{(2)}(t_0)=0$. The scalar particle momentum
distribution can be calculated by using the coefficient
$c_{\bf{k}}^{(2)}(t)$ as
 \begin{eqnarray}\label{eq:s27}
f_{\bf k}(t)=2|c_{\bf{k}}^{(2)}(t)|^2,
\end{eqnarray}
with
\begin{align}\label{eq:s28}
\Omega_{\mathbf{k}}(t) &=\sqrt{\mathcal{E}^2(t) - \frac{1}{2}
\dot{P}(t) - \frac{1}{4} P^2(t)}.
\end{align}

If only gravity is present but without any external electric
field, then Eq. \eqref{eq:s25} reduces to the exact GQTLM of the
traditional Hawking radiation. If only an external electric field
is present but without any gravity, then Eq. \eqref{eq:s25}
reduces to the GQTLM of the Schwinger effect~\cite{Amat:2024nvg}.
If both the gravity and the external electric field can not be
ignored, then Eq. \eqref{eq:s25} corresponds to the GQTLM that
considers the coupling of the Hawking radiation and the Schwinger
effect. Especially, note that the difference between Eq.
\eqref{eq:s25} and Eq. \eqref{eq:s21} is that the former does not
include the inflationary effect and the interaction of the
electromagnetic field with the inflation.

\section{GQTLM in context of the Schwinger effect}
\label{sec:c}

For compact astrophysical objects, such as neutron stars or
strange quark stars, the scalar QED Lagrangian density under
time-dependent curved spacetime and electric field is
\begin{align}
\mathcal{L}_{3}&=\sqrt{-g}\left(g^{\mu\nu}(\mathcal{D}_{\mu}\chi)^{\dagger}(\mathcal{D}_{\nu}\chi)-(m^{2}-\xi_{3} R_{3})\chi^{\dagger}\chi -\frac{1}{4}F_{\mu\nu}F^{\mu\nu} \right).
\end{align}
Similarly, the exact GQTLM can be established by using this
Lagrangian. For simplicity, we do not repeat the process here.

\section{GQTLM in context of the gravitational pair production}
\label{sec:d}

When a massive compact astrophysical object merge with a black
hole, the varying curvature of spacetime caused by the strong
gravity of the two objects, along with the strong electric field
of the charging black hole, may lead to the quantum vacuum decay.
Particle-antiparticle pairs would be produced during the process.
Again, GQTLM involving time-dependent curved spacetime and
electric field is needed. In this case, the scalar QED Lagrangian
density can be expressed as
\begin{align}
\mathcal{L}_{4}&=\sqrt{-g}\left(g^{\mu\nu}(\mathcal{D}_{\mu}\chi)^{\dagger}(\mathcal{D}_{\nu}\chi)-(m^{2}-\xi_{4} R_{4})\chi^{\dagger}\chi -\frac{1}{4}F_{\mu\nu}F^{\mu\nu} \right).
\end{align}
Following a similar derivation, the exact GQTLM can be established
in context of the gravitational pair production.

\section{Quantum kinetic theory for curved spacetime with electric field}
\label{sec:e}

The quantum Vlasov equation (QVE) and the Wigner function are
collectively called Quantum kinetic theory (QKT). QVE can also be
used to deal with scalar vacuum pair production for the early
universe, Hawking radiation, Schwinger effect and gravitational
pair production. This can be achieved by using two coefficients,
$\alpha_{\bf{k}}(t)$ and $\beta_{\bf{k}}(t)$, because there are
only three real degrees of freedom in the scalar vacuum pair
production. As a result, we have
\begin{eqnarray}
\mathcal{F}_{a,~\mathbf{k}}(t)&=&\big|\beta_{a,~\mathbf{k}}\big|^{2},\label{eq:s31}\\
\mathcal{G}_{a,~\mathbf{k}}(t)&=&-2\Re e\left(\alpha_{a,~\mathbf{k}}~\beta^{*}_{a,~\mathbf{k}} e^{-2i\int^t d\tau \Omega_{a,~\mathbf{k}}(\tau)}\right),\label{eq:s32}\\
\mathcal{H}_{a,~\mathbf{k}}(t)&=&2\Im m\left(\alpha_{a,~\mathbf{k}}~\beta^{*}_{a,~\mathbf{k}} e^{-2i\int^t d\tau \Omega_{a,~\mathbf{k}}(\tau)}\right),\label{eq:s33}
\end{eqnarray}
where the index $a=\{1, 2, 3, 4\}$ represents the four situations
of the early universe, Hawking radiation, the Schwinger effect,
and gravitational pair production, respectively. Taking the time
derivative of the above set of equations, we have
\begin{eqnarray}
\dot{\mathcal{F}}_{a,~\mathbf{k}}&=&\frac{1}{2} W_{a,~\bf{k}}(t)~\mathcal{G}_{a,~\mathbf{k}},\label{eq:s34}\\
\dot{\mathcal{G}}_{a,~\mathbf{k}}&=&W_{a,~\bf{k}}(t)~(1+2\mathcal{F}_{a,~\mathbf{k}})-2\Omega_{a,~\mathbf{k}}~\mathcal{H}_{a,~\mathbf{k}},\label{eq:s35}\\
\dot{\mathcal{H}}_{a,~\mathbf{k}}&=&2\Omega_{a,~\mathbf{k}}~\mathcal{G}_{a,~\mathbf{k}},
\label{eq:s36}
\end{eqnarray}
where
\begin{align}\label{eq:s37}
W_{a,~\bf{k}}(t)=\frac{1}{2} Q_{a,~\bf{k}}(t).
\end{align}
The initial condition is given by
\begin{equation}\label{eq:s38}
\mathcal{F}_{a,~\mathbf{k}} (t_0)=\mathcal{G}_{a,~\mathbf{k}} (t_0)=\mathcal{H}_{a,~\mathbf{k}} (t_0)=0.
\end{equation}

The scalar particle momentum distribution is expressed as
\begin{eqnarray}\label{eq:s39}
f_{a,~\bf k}(t)=\mathcal{F}_{a,~\mathbf{k}} (t).
\end{eqnarray}
The particle number density can be calculated by integrating the
momentum distribution function $\mathcal{F}_{a,~\mathbf{k}} (t)$
over the momentum $\mathbf{k}$,
\begin{equation}\label{eq:s40}
n_{a}(t)=\int \frac{d^3k}{(2\pi)^3}f_{a,~\bf k}(t).
\end{equation}
It can be seen that our exact GQTLM is equivalent to the QVE in
cases of the early universe, Hawking radiation, the Schwinger
effect, and gravitational pair production. When the spacetime
curvature is weak so that it could be neglected, our Eqs.
\eqref{eq:s34}-\eqref{eq:s36} will become consistent with the
results presented in Ref.~\cite{Li:2019rex}, which is a natural
expectation.


\begin{thebibliography}{apsrev4-1}

\bibitem{Piazza:2001}
A. Di Piazza, E. Fiordilino, and M.H. Mittleman,
Analytical study of the spectrum emitted by a two level atom driven by a strong electric field,
\href{https://journals.aps.org/pra/pdf/10.1103/PhysRevA.64.013414}{Phys. Rev. A \textbf{64}, 013414 (2001)}.

\bibitem{Avetissian:2002ucr}
H.~K.~Avetissian, A.~K.~Avetissian, G.~F.~Mkrtchian, and K.~V.~Sedrakian,
Electron-positron pair production in the field of superstrong oppositely directed laser beams,
\href{https://journals.aps.org/pre/pdf/10.1103/PhysRevE.66.016502}{Phys. Rev. E \textbf{66}, 016502 (2002)}.

\bibitem{Akkermans:2011yn}
E.~Akkermans and G.~V.~Dunne,
Ramsey Fringes and Time-domain Multiple-Slit Interference from Vacuum,
\href{https://journals.aps.org/prl/pdf/10.1103/PhysRevLett.108.030401}{Phys. Rev. Lett. \textbf{108}, 030401 (2012)}.

\bibitem{Kaminski:2018ywj}
J.~Z.~Kami\'nski, M.~Twardy, and K.~Krajewska,
Diffraction at a time grating in electron-positron pair creation from the vacuum,
\href{https://journals.aps.org/prd/pdf/10.1103/PhysRevD.98.056009}{Phys. Rev. D \textbf{98}, 056009 (2018)}.

\bibitem{Krajewska:2019vqd}
K.~Krajewska and J.~Z.~Kami\'nski,
Unitary versus pseudounitary time evolution and statistical effects in the dynamical Sauter-Schwinger process,
\href{https://journals.aps.org/pra/pdf/10.1103/PhysRevA.100.062116}{Phys. Rev. A \textbf{100}, 062116 (2019)}.

\bibitem{Fiordilino:2021zkp}
E.~Fiordilino,
Quest for time variation of Planck constant: A new time standard and parametric chaos,
\href{https://link.springer.com/content/pdf/10.1140/epjp/s13360-020-01031-1}{Eur. Phys. J. Plus \textbf{136}, 54 (2021)}.

\bibitem{Dunne:2022zlx}
G.~V.~Dunne, A.~Florio, and D.~E.~Kharzeev,
Entropy suppression through quantum interference in electric pulses,
\href{https://journals.aps.org/prd/pdf/10.1103/PhysRevD.108.L031901}{Phys. Rev. D \textbf{108}, L031901 (2023)}.

\bibitem{Bechler:2023kjx}
A.~Bechler, F.~Cajiao V\'elez, K.~Krajewska, and J.~Z.~Kami\'nski,
Vortex Structures and Momentum Sharing in Dynamic Sauter-Schwinger Process,
\href{https://appol.ifpan.edu.pl/index.php/appa/article/view/143_s18/143_s18}{Acta Phys. Polon. A \textbf{143}, S18 (2023)}.

\bibitem{Keldysh:1965ojf}
L.~V.~Keldysh,
Ionization in the Field of a Strong Electromagnetic Wave,
\href{https://inspirehep.net/files/6697e05d52e411291acc8238a780db45}{J. Exp. Theor. Phys. \textbf{20}, 1307 (1965)}.

\bibitem{2005PhRvL}
F.~{Lindner} {\em et~al.},
Attosecond Double-Slit Experiment,
\href{https://journals.aps.org/prl/pdf/10.1103/PhysRevLett.95.040401}{Phys. Rev. Lett. \textbf{95}, 040401 (2005)}.

\bibitem{Krausz:2009zz}
F.~Krausz and M.~Ivanov,
Attosecond physics,
\href{https://journals.aps.org/rmp/pdf/10.1103/RevModPhys.81.163}{Rev. Mod. Phys. \textbf{81}, 163 (2009)}.

\bibitem{Keski-Vakkuri:1996lbi}
E.~Keski-Vakkuri and P.~Kraus,
Tunneling in a time dependent setting,
\href{https://journals.aps.org/prd/pdf/10.1103/PhysRevD.54.7407}{Phys. Rev. D \textbf{54}, 7407 (1996)}.

\bibitem{zueco2008landau}
D.~Zueco, P.~H{\"a}nggi, and S.~Kohler,
Landau-Zener tunnelling in dissipative circuit QED,
\href{https://beta.iopscience.iop.org/article/10.1088/1367-2630/10/11/115012/pdf}{New Journal of Physics {\bf 10}, 115012 (2008)}.

\bibitem{oka2009nonequilibrium}
T.~Oka and H.~Aoki,
\newblock Nonequilibrium quantum breakdown in a strongly correlated electron system,
\newblock in {\em Quantum and Semi-classical Percolation and Breakdown in Disordered Solids}, pp. 1--35, Springer, 2009.

\bibitem{Shevchenko:2010ms}
S.~N.~Shevchenko, S.~Ashhab, and F.~Nori,
Landau-Zener-Stuckelberg Interferometry,
\href{https://www.sciencedirect.com/science/article/pii/S0370157310000815?ref=pdf_download&fr=RR-9&rr=8e8868632fc81055}{Phys. Rept. \textbf{492}, 1 (2010)}.

\bibitem{li2010carrier}
F.~{Lindner} {\em et~al.},
Carrier-Envelope Phase Effect on Atomic Excitation by Few-Cycle rf Pulses,
\href{https://journals.aps.org/prl/pdf/10.1103/PhysRevLett.104.103001}{Phys. Rev. Lett. \textbf{104}, 103001 (2010)}.

\bibitem{jha2011experimental}
P.~K. Jha, Y.~V. Rostovtsev, H.~Li, V.~A. Sautenkov, and M.~O. Scully,
Experimental observation of carrier-envelope-phase effects by multicycle pulse,
\href{https://journals.aps.org/pra/pdf/10.1103/PhysRevA.83.033404}{Phys. Rev. A \textbf{83}, 033404 (2011)}.

\bibitem{miller1968semiclassical}
W. H. Miller,
Semiclassical Treatment of Multiple Turning-Point Problems-Phase Shifts and Eigenvalues,
\href{https://pubs.aip.org/aip/jcp/article/48/4/1651/83744/Semiclassical-Treatment-of-Multiple-Turning-Point}{J. Chem. Phys. \textbf{48}, 1651 (1968)}.

\bibitem{saha2011tunneling}
R.~Saha and V.~S. Batista,
Tunneling under Coherent Control by Sequences of Unitary Pulses,
\href{https://pubs.acs.org/doi/epdf/10.1021/jp108331x?ref=article_openPDF}{J. Phys. Chem. B {\bf 115}, 5234 (2011)}.

\bibitem{Brout:1995rd}
R.~Brout, S.~Massar, R.~Parentani, and P.~Spindel,
A Primer for black hole quantum physics,
\href{https://www.sciencedirect.com/science/article/pii/0370157395000085}{Phys. Rept. \textbf{260}, 329 (1995)}.

\bibitem{Parikh:1999mf}
M.~K.~Parikh and F.~Wilczek,
Hawking radiation as tunneling,
\href{https://journals.aps.org/prl/pdf/10.1103/PhysRevLett.85.5042}{Phys. Rev. Lett. \textbf{85}, 5042 (2000)}.

\bibitem{Parker:1968mv}
L.~Parker,
Particle creation in expanding universes,
\href{https://journals.aps.org/prl/pdf/10.1103/PhysRevLett.21.562}{Phys. Rev. Lett. \textbf{21}, 562 (1968)}.

\bibitem{Greiner1985-qt}
W.~Greiner, B.~M{\"u}ller, and J.~Rafelski,
\newblock {\em Quantum electrodynamics of strong fields} (Springer, Berlin, Germany, 1985).

\bibitem{Blaschke:2017igl}
D.~B. Blaschke, S.~A. Smolyansky, A.~Panferov, and L.~Juchnowski,
\newblock {Particle Production in Strong Time-dependent Fields},
\newblock in {\em {Quantum Field Theory at the Limits}: {from Strong Fields to Heavy Quarks}}, pp. 1--23, (2017).

\bibitem{Klich:2008un}
I.~Klich and L.~Levitov,
Quantum Noise as an Entanglement Meter,
\href{https://journals.aps.org/prl/pdf/10.1103/PhysRevLett.102.100502}{Phys. Rev. Lett. \textbf{102}, 100502 (2009)}.

\bibitem{Goren}
T.~Goren, K.~L. Hur, and E.~Akkermans,
Ramsey interferometry of particle-hole pairs in tunnel junctions,
\href{https://arxiv.org/pdf/1611.06738}{[arXiv:1611.06738 [quant-ph]]}.

\bibitem{Jaekel:1997hr}
M.~T.~Jaekel and S.~Reynaud,
Movement and fluctuations of the vacuum,
\href{https://iopscience.iop.org/article/10.1088/0034-4885/60/9/001/pdf}{Rept. Prog. Phys. \textbf{60}, 863 (1997)}.

\bibitem{Dodonov:2010zza}
V.~V.~Dodonov,
Current status of the dynamical Casimir effect,
\href{https://iopscience.iop.org/article/10.1088/0031-8949/82/03/038105/pdf}{Phys. Scripta \textbf{82}, 038105 (2010)}.

\bibitem{Majczak:2024hmt}
M.~M.~Majczak, K.~Krajewska, J.~Z.~Kami\'nski, and A.~Bechler,
Scattering matrix approach to dynamical Sauter-Schwinger process: Spin- and helicity-resolved momentum distributions,
\href{https://arxiv.org/pdf/2403.15206}{[arXiv:2403.15206 [quant-ph]]}.

\bibitem{Solinas:2020woq}
P.~Solinas, A.~Amoretti and F.~Giazotto,
Sauter-Schwinger effect in a Bardeen-Cooper-Schrieffer superconductor,
\href{https://journals.aps.org/prl/pdf/10.1103/PhysRevLett.126.117001}{Phys. Rev. Lett. \textbf{126}, 117001 (2021)}.

\bibitem{Boskovic:2024fga}
M.~Bo\v{s}kovi\'c, M.~Koschnitzke, and R.~A.~Porto,
Signatures of Ultralight Bosons in the Orbital Eccentricity of Binary Black Holes,
\href{https://journals.aps.org/prl/pdf/10.1103/PhysRevLett.133.121401}{Phys. Rev. Lett. \textbf{133}, 12 (2024)}.

\bibitem{Tomaselli:2024faa}
G.~M.~Tomaselli,
Gravitational atoms and black hole binaries,
Ph.D. thesis, University of Amsterdam,
\href{https://pure.uva.nl/ws/files/196138381/Thesis.pdf}{https://pure.uva.nl/ws/files/196138381/Thesis.pdf}.

\bibitem{Huang:1970iq}
K.~Huang and S.~Weinberg,
Ultimate temperature and the early universe,
\href{https://journals.aps.org/prl/pdf/10.1103/PhysRevLett.25.895}{Phys. Rev. Lett. \textbf{25}, 895 (1970)}.

\bibitem{Hawking:1974rv}
S.~W.~Hawking,
Black hole explosions,
\href{https://www.nature.com/articles/248030a0}{Nature \textbf{248}, 30 (1974)}.

\bibitem{Hawking:1975vcx}
S.~W.~Hawking,
Particle Creation by Black Holes,
\href{https://link.springer.com/article/10.1007/BF02345020}{Commun. Math. Phys. \textbf{43}, 199 (1975)}.

\bibitem{Gibbons:1976ue}
G.~W.~Gibbons and S.~W.~Hawking,
Action Integrals and Partition Functions in Quantum Gravity,
\href{https://journals.aps.org/prd/pdf/10.1103/PhysRevD.15.2752}{Phys. Rev. D \textbf{15}, 2752 (1977)}.

\bibitem{Dumlu:2020wvd}
C.~K.~Dumlu,
Stokes phenomenon and Hawking radiation,
\href{https://journals.aps.org/prd/pdf/10.1103/PhysRevD.102.125006}{Phys. Rev. D \textbf{102}, 125006 (2020)}.

\bibitem{Trevisan:2024jvn}
S.~Trevisan, F.~Belgiorno and S.~L.~Cacciatori,
Exact solutions for analog Hawking effect in dielectric media,
\href{https://journals.aps.org/prd/pdf/10.1103/PhysRevD.110.085009}{Phys. Rev. D \textbf{110}, 085009 (2024)}.

\bibitem{Schwinger:1951nm}
J.~S.~Schwinger,
On gauge invariance and vacuum polarization,
\href{https://journals.aps.org/pr/pdf/10.1103/PhysRev.82.664}{Phys. Rev. \textbf{82}, 664 (1951)}.

\bibitem{Bialynicki-Birula:1991jwl}
I.~Bialynicki-Birula, P.~Gornicki, and J.~Rafelski,
Phase space structure of the Dirac vacuum,
\href{https://journals.aps.org/prd/pdf/10.1103/PhysRevD.44.1825}{Phys. Rev. D \textbf{44}, 1825 (1991)}.

\bibitem{Alkofer:2001ik}
R.~Alkofer, M.~B.~Hecht, C.~D.~Roberts, S.~M.~Schmidt and D.~V.~Vinnik,
Pair creation and an x-ray free electron laser,
\href{https://journals.aps.org/prl/pdf/10.1103/PhysRevLett.87.193902}{Phys. Rev. Lett. \textbf{87}, 193902 (2001)}.

\bibitem{Hebenstreit:2011wk}
F.~Hebenstreit, R.~Alkofer, and H.~Gies,
Particle self-bunching in the Schwinger effect in spacetime-dependent electric fields,
\href{https://journals.aps.org/prl/pdf/10.1103/PhysRevLett.107.180403}{Phys. Rev. Lett. \textbf{107}, 180403 (2011)}.

\bibitem{Amat:2022uxq}
O.~Amat, L.~N.~Hu, A.~Sawut, M.~Mohamedsedik, M.~A.~Bake, and B.~S.~Xie,
Schwinger pair production rate and time for some space-dependent fields via worldline instantons formalism,
\href{https://link.springer.com/article/10.1140/epjd/s10053-022-00519-y}{Eur. Phys. J. D \textbf{76}, 188 (2022)}.

\bibitem{Kohlfurst:2021skr}
C.~Kohlf\"urst, N.~Ahmadiniaz, J.~Oertel, and R.~Sch\"utzhold,
Sauter-Schwinger Effect for Colliding Laser Pulses,
\href{https://journals.aps.org/prl/pdf/10.1103/PhysRevLett.129.241801}{Phys. Rev. Lett. \textbf{129}, 241801 (2022)}.

\bibitem{Ilderton:2021zej}
A.~Ilderton,
Physics of adiabatic particle number in the Schwinger effect,
\href{https://journals.aps.org/prd/pdf/10.1103/PhysRevD.105.016021}{Phys. Rev. D \textbf{105}, 016021 (2022)}.

\bibitem{Amat:2023vwv}
O.~Amat, L.~N.~Hu, M.~A.~Bake, M.~Mohamedsedik, and B.~S.~Xie,
Effect of spatially oscillating field on Schwinger pair production,
\href{https://journals.aps.org/prd/pdf/10.1103/PhysRevD.108.056011}{Phys. Rev. D \textbf{108}, 056011 (2023)}.

\bibitem{Aleksandrov:2024rsz}
I.~A.~Aleksandrov, A.~Kudlis and A.~I.~Klochai,
Kinetic theory of vacuum pair production in uniform electric fields revisited,
\href{https://journals.aps.org/prresearch/pdf/10.1103/PhysRevResearch.6.043009}{Phys. Rev. Res. \textbf{6}, 043009 (2024)}.

\bibitem{Litim:2007iu}
D.~F.~Litim and T.~Plehn,
Signatures of gravitational fixed points at the LHC,
\href{https://journals.aps.org/prl/pdf/10.1103/PhysRevLett.100.131301}{Phys. Rev. Lett. \textbf{100}, 131301 (2008)}.

\bibitem{Wondrak:2023zdi}
M.~F.~Wondrak, W.~D.~van Suijlekom, and H.~Falcke,
Gravitational Pair Production and Black Hole Evaporation,
\href{https://journals.aps.org/prl/pdf/10.1103/PhysRevLett.130.221502}{Phys. Rev. Lett. \textbf{130}, 221502 (2023)}.

\bibitem{Hebenstreit:2009km}
F.~Hebenstreit, R.~Alkofer, G.~V.~Dunne and H.~Gies,
Momentum signatures for Schwinger pair production in short laser pulses with sub-cycle structure,
\href{https://journals.aps.org/prl/pdf/10.1103/PhysRevLett.102.150404}{Phys. Rev. Lett. \textbf{102}, 150404 (2009)}.

\bibitem{Dumlu:2010ua}
C.~K.~Dumlu and G.~V.~Dunne,
The Stokes Phenomenon and Schwinger Vacuum Pair Production in Time-Dependent Laser Pulses,
\href{https://journals.aps.org/prl/pdf/10.1103/PhysRevLett.104.250402}{Phys. Rev. Lett. \textbf{104}, 250402 (2010)}.

\bibitem{Enomoto:2022mti}
S.~Enomoto and T.~Matsuda,
The Exact WKB analysis and the Stokes phenomena of the Unruh effect and Hawking radiation,
\href{https://link.springer.com/article/10.1007/JHEP12(2022)037}{JHEP \textbf{12}, 037 (2022)}.

\bibitem{Amat:2024nvg}
O.~Amat, H.~H.~Fan, S.~Tang, Y.~F.~Huang and B.~S.~Xie,
Spin resolved momentum spectra for vacuum pair production via a generalized two level model,
\href{https://arxiv.org/pdf/2409.11833}{[arXiv:2409.11833 [hep-ph]]}.

\bibitem{Amat:2025zep}
O.~Amat, N.~Nurmamat, Y.~F.~Huang, C.~M.~Li, J.~J.~Geng, C.~R.~Hu, Z.~C.~Zou, X.~F.~Dong, C.~Deng and F.~Xu, \textit{et al.}
Axial current as the origin of quantum intrinsic orbital angular momentum,
\href{https://arxiv.org/pdf/2502.06156}{[arXiv:2502.06156 [hep-ph]]}.

\bibitem{Fedotov:2022ely}
A.~Fedotov, A.~Ilderton, F.~Karbstein, B.~King, D.~Seipt, H.~Taya, and G.~Torgrimsson,
Advances in QED with intense background fields,
\href{https://www.sciencedirect.com/science/article/pii/S0370157323000352?via%3Dihub}{Phys. Rept. \textbf{1010}, 1 (2023)}.

\bibitem{Allen:1992zz}
L.~Allen, M.~W.~Beijersbergen, R.~J.~C.~Spreeuw, and J.~P.~Woerdman,
Orbital angular momentum of light and the transformation of Laguerre-Gaussian laser modes,
\href{https://journals.aps.org/pra/pdf/10.1103/PhysRevA.45.8185}{Phys. Rev. A \textbf{45}, 8185 (1992)}.

\bibitem{Mair:2001fyp}
A.~Mair, A.~Vaziri, G.~Weihs and A.~Zeilinger,
Entanglement of the orbital angular momentum states of photons,
\href{https://www.nature.com/articles/35085529}{Nature \textbf{412}, 313 (2001)}.

\bibitem{Jentschura:2010ap}
U.~D.~Jentschura and V.~G.~Serbo,
Generation of High-Energy Photons with Large Orbital Angular Momentum by Compton Backscattering,
\href{https://journals.aps.org/prl/pdf/10.1103/PhysRevLett.106.013001}{Phys. Rev. Lett. \textbf{106}, 013001 (2011)}.

\bibitem{PhysRevLett.107.174802}
K.~Y.~Bliokh, M.~R.~Dennis,and F.~Nori
Relativistic Electron Vortex Beams: Angular Momentum and Spin-Orbit Interaction,
\href{https://journals.aps.org/prl/pdf/10.1103/PhysRevLett.107.174802}{Phys. Rev. Lett. \textbf{107}, 174802 (2011)}.

\bibitem{Bahrdt:2013eoa}
J.~Bahrdt, K.~Holldack, P.~Kuske, R.~M\"uller, M.~Scheer, and P.~Schmid,
First Observation of Photons Carrying Orbital Angular Momentum in Undulator Radiation,
\href{https://journals.aps.org/prl/pdf/10.1103/PhysRevLett.111.034801}{Phys. Rev. Lett. \textbf{111}, 034801 (2013)}.

\bibitem{Bliokh:2015yhi}
K.~Y.~Bliokh, F.~J.~Rodr\'\i{}guez-Fortu\~no, F.~Nori, and A.~V.~Zayats,
Spin\textendash{}orbit interactions of light,
\href{https://www.nature.com/articles/nphoton.2015.201}{Nature Photon. \textbf{9}, 796 (2015)}.

\bibitem{RevModPhys.89.035004}
S.~M.~Lloyd, M.~Babiker, G.~Thirunavukkarasu, and J.~Yuan
Electron vortices: Beams with orbital angular momentum,
\href{https://journals.aps.org/rmp/pdf/10.1103/RevModPhys.89.035004}{Rev. Mod. Phys. \textbf{89}, 035004 (2017)}.

\bibitem{Bliokh:2017uvr}
K.~Y.~Bliokh, I.~P.~Ivanov, G.~Guzzinati, L.~Clark, R.~Van Boxem, A.~B\'ech\'e, R.~Juchtmans, M.~A.~Alonso, P.~Schattschneider, and F.~Nori, \textit{et al.}
Theory and applications of free-electron vortex states,
\href{https://www.sciencedirect.com/science/article/pii/S0370157317301515?via%3Dihub}{Phys. Rept. \textbf{690}, 1 (2017)}.

\bibitem{Ivanov:2019vxe}
I.~P.~Ivanov, N.~Korchagin, A.~Pimikov, and P.~Zhang,
Doing spin physics with unpolarized particles,
\href{https://journals.aps.org/prl/pdf/10.1103/PhysRevLett.124.192001}{Phys. Rev. Lett. \textbf{124}, 192001 (2020)}.

\bibitem{Ivanov:2022jzh}
I.~P.~Ivanov,
Promises and challenges of high-energy vortex states collisions,
\href{https://www.sciencedirect.com/science/article/pii/S0146641022000461?via%3Dihub}{Prog. Part. Nucl. Phys. \textbf{127}, 103987 (2022)}.

\bibitem{Karlovets:2022evc}
D.~V.~Karlovets, S.~S.~Baturin, G.~Geloni, G.~K.~Sizykh, and V.~G.~Serbo,
Generation of vortex particles via generalized measurements,
\href{https://link.springer.com/article/10.1140/epjc/s10052-022-10991-w}{Eur. Phys. J. C \textbf{82}, 1008 (2022)}.

\bibitem{Lu:2023wrf}
Z.~W.~Lu, L.~Guo, Z.~Z.~Li, M.~Ababekri, F.~Q.~Chen, C.~Fu, C.~Lv, R.~Xu, X.~Kong and Y.~F.~Niu, \textit{et al.}
Manipulation of Giant Multipole Resonances via Vortex \ensuremath{\gamma} Photons,
\href{https://journals.aps.org/prl/pdf/10.1103/PhysRevLett.131.202502}{Phys. Rev. Lett. \textbf{131}, 202502 (2023)}.

\bibitem{Ababekri:2022mob}
M.~Ababekri, R.~T.~Guo, F.~Wan, B.~Qiao, Z.~Li, C.~Lv, B.~Zhang, W.~Zhou, Y.~Gu, and J.~X.~Li,
Vortex \ensuremath{\gamma} photon generation via spin-to-orbital angular momentum transfer in nonlinear Compton scattering,
\href{https://journals.aps.org/prd/pdf/10.1103/PhysRevD.109.016005}{Phys. Rev. D \textbf{109}, 016005 (2024)}.

\bibitem{Ababekri:2024cyd}
M.~Ababekri, J.~L.~Zhou, R.~T.~Guo, Y.~Z.~Ren, Y.~H.~Kou, Q.~Zhao, Z.~P.~Li, and J.~X.~Li,
Generation of ultrarelativistic vortex leptons with large orbital angular momenta,
\href{https://journals.aps.org/prd/pdf/10.1103/PhysRevD.110.076024}{Phys. Rev. D \textbf{110}, 076024 (2024)}.

\bibitem{Ababekri:2024glc}
M.~Ababekri, Y.~Wang, R.~T.~Guo, Z.~P.~Li, and J.~X.~Li,
Dynamics of relativistic vortex electrons in external laser fields,
\href{https://journals.aps.org/pra/pdf/10.1103/PhysRevA.110.052207}{Phys. Rev. A \textbf{110}, 052207 (2024)}.

\bibitem{Session:2023hoq}
D.~Session, M.~J.~Mehrabad, N.~Paithankar, T.~Grass, C.~J.~Eckhardt, B.~Cao, D.~G.~S.~Forero, K.~Li, M.~S.~Alam, and K.~Watanabe, \textit{et al.}
Optical pumping of electronic quantum Hall states with vortex light,
\href{https://doi.org/10.1038/s41566-024-01565-1}{Nature Photon. (2024)}.

\bibitem{Kohlfurst:2013ura}
C.~Kohlf\"urst, H.~Gies and R.~Alkofer,
Effective mass signatures in multiphoton pair production,
\href{https://journals.aps.org/prl/pdf/10.1103/PhysRevLett.112.050402}{Phys. Rev. Lett. \textbf{112}, 050402 (2014)}.

\bibitem{Gorbar:2019fpj}
E.~V.~Gorbar, A.~I.~Momot, O.~O.~Sobol, and S.~I.~Vilchinskii,
Kinetic approach to the Schwinger effect during inflation,
\href{https://journals.aps.org/prd/pdf/10.1103/PhysRevD.100.123502}{Phys. Rev. D \textbf{100}, 123502 (2019)}.

\bibitem{Friedmann:1924bb}
A.~Friedmann,
On the Possibility of a world with constant negative curvature of space,
\href{https://link.springer.com/article/10.1007/BF01328280}{Z. Phys. \textbf{21}, 326 (1924)}.

\bibitem{Robertson:1929zz}
H.~P.~Robertson,
The Uncertainty Principle,
\href{https://journals.aps.org/pr/pdf/10.1103/PhysRev.34.163}{Phys. Rev. \textbf{34}, 163 (1929)}.

\bibitem{Lemaitre:1931zza}
A.~G.~Lema\^itre,
A Homogeneous Universe of Constant Mass and Increasing Radius accounting for the Radial Velocity of Extra-galactic Nebul\ae{},
\href{https://academic.oup.com/mnras/article/91/5/483/985165?login=true}{Mon. Not. Roy. Astron. Soc. \textbf{91}, 483 (1931)}.

\bibitem{Robertson:1933zz}
H.~P.~Robertson,
Relativistic Cosmology,
\href{https://journals.aps.org/rmp/pdf/10.1103/RevModPhys.5.62}{Rev. Mod. Phys. \textbf{5}, 62 (1933)}.

\bibitem{Walker:1937qxv}
A.~G.~Walker,
On Milne's Theory of World-Structure,
\href{https://londmathsoc.onlinelibrary.wiley.com/doi/abs/10.1112/plms/s2-42.1.90}{Proc. Lond. Math. Soc. s \textbf{2-42}, 90 (1937)}.

\bibitem{Weinberg:1988cp}
S.~Weinberg,
The Cosmological Constant Problem,
\href{https://journals.aps.org/rmp/pdf/10.1103/RevModPhys.61.1}{Rev. Mod. Phys. \textbf{61}, 1 (1989)}.

\bibitem{Pettinari:2016}
G.~W.~Pettinari,
The Standard Model of Cosmology. In: The Intrinsic Bispectrum of the Cosmic Microwave Background,
Ph.D. thesis, Springer Theses,
\href{https://link.springer.com/chapter/10.1007/978-3-319-21882-3_2}{https://link.springer.com/chapter/10.1007/978-3-319-21882-3\_2}.

\bibitem{Kitamoto:2020tjm}
H.~Kitamoto,
No-go theorem of anisotropic inflation via Schwinger mechanism,
\href{https://journals.aps.org/prd/pdf/10.1103/PhysRevD.103.063521}{Phys. Rev. D \textbf{103}, 063521 (2021)}.

\bibitem{Okano:2021pva}
S.~Okano and T.~Fujita,
When does the Schwinger preheating occur?,
\href{https://iopscience.iop.org/article/10.1088/1475-7516/2022/03/040/pdf}{JCAP \textbf{03}, 040 (2022)}.

\bibitem{Meimanat:2023hjq}
O.~G.~Meimanat and E.~Bavarsad,
Induced energy-momentum tensor in de~Sitter scalar QED and its implication for induced currents,
\href{https://journals.aps.org/prd/pdf/10.1103/PhysRevD.107.125001}{Phys. Rev. D \textbf{107}, 125001 (2023)}.

\bibitem{Lysenko:2023wrs}
A.~V.~Lysenko and O.~O.~Sobol,
Quantum kinetic approach to the Schwinger production of scalar particles in an expanding universe,
\href{https://link.springer.com/article/10.1007/s10714-024-03226-8}{Gen. Rel. Grav. \textbf{56}, 39 (2024)}.

\bibitem{Kolb:1981hk}
E.~W.~Kolb and M.~S.~Turner,
The Early Universe,
\href{https://www.nature.com/articles/294521a0}{Nature \textbf{294}, 521 (1981)}.

\bibitem{Guth:2005zr}
A.~H.~Guth and D.~I.~Kaiser,
Inflationary cosmology: Exploring the Universe from the smallest to the largest scales,
\href{https://www.science.org/doi/epdf/10.1126/science.1107483}{Science \textbf{307}, 884 (2005)}.

\bibitem{Podolsky:2005bw}
D.~I.~Podolsky, G.~N.~Felder, L.~Kofman, and M.~Peloso,
Equation of state and beginning of thermalization after preheating,
\href{https://journals.aps.org/prd/pdf/10.1103/PhysRevD.73.023501}{Phys. Rev. D \textbf{73}, 023501 (2006)}.

\bibitem{Shakeri:2019mnt}
S.~Shakeri, M.~A.~Gorji and H.~Firouzjahi,
Schwinger Mechanism During Inflation,
\href{https://journals.aps.org/prd/pdf/10.1103/PhysRevD.99.103525}{Phys. Rev. D \textbf{99}, 103525 (2019)}.

\bibitem{Tryon:1973xi}
E.~P.~Tryon,
Is the universe a vacuum fluctuation,
\href{https://www.nature.com/articles/246396a0}{Nature \textbf{246}, 396 (1973)}.

\bibitem{PoncedeLeon:2017usu}
J.~Ponce de Leon,
Regular Reissner-Nordstr\"om black hole solutions from linear electrodynamics,
\href{https://journals.aps.org/prd/pdf/10.1103/PhysRevD.95.124015}{Phys. Rev. D \textbf{95}, 124015 (2017)}.

\bibitem{Balart:2014cga}
L.~Balart and E.~C.~Vagenas,
Regular black holes with a nonlinear electrodynamics source,
\href{https://journals.aps.org/prd/pdf/10.1103/PhysRevD.90.124045}{Phys. Rev. D \textbf{90}, 124045 (2014)}.

\bibitem{Prakapenia:2023fvx}
M.~Prakapenia and G.~Vereshchagin,
Pair Creation in Hot Electrosphere of Compact Astrophysical Objects,
\href{https://iopscience.iop.org/article/10.3847/1538-4357/ad24ee/pdf}{Astrophys. J. \textbf{963}, 149 (2024)}.

\bibitem{Geng:2015uja}
J.~J.~Geng, Y.~F.~Huang, and T.~Lu,
Coalescence of Strange-quark Planets With Strange Stars: a new Kind of Source for Gravitational Wave Bursts,
\href{https://iopscience.iop.org/article/10.1088/0004-637X/804/1/21/pdf}{Astrophys. J. \textbf{804}, 21 (2015)}.

\bibitem{Zhang:2024uco}
X.~L.~Zhang, Z.~C.~Zou, Y.~F.~Huang, H.~X.~Gao, P.~Wang, L.~Cui, and X.~Liu,
Gravitational wave emission from close-in strange quark planets around strange stars with magnetic interactions,
\href{https://academic.oup.com/mnras/article/531/4/3905/7688456?login=true}{Mon. Not. Roy. Astron. Soc. \textbf{531}, 3905 (2024)}.

\bibitem{Zou:2022wtp}
Z.~C.~Zou and Y.~F.~Huang,
Gravitational-wave Emission from a Primordial Black Hole Inspiraling inside a Compact Star: A Novel Probe for Dense Matter Equation of State,
\href{https://iopscience.iop.org/article/10.3847/2041-8213/ac5ea6/pdf}{Astrophys. J. Lett. \textbf{928}, L13 (2022)}.

\bibitem{Zou:2022voz}
Z.~C.~Zou, Y.~F.~Huang, and X.~L.~Zhang,
Gravitational Waves from Strange Star Core\textendash{}Crust Oscillation,
\href{https://www.mdpi.com/2218-1997/8/9/442}{Universe \textbf{8}, 442 (2022)}.

\bibitem{Prakapenia:2023tsw}
M.~Prakapenia and G.~Vereshchagin,
Pauli blocking effects on pair creation in strong electric field,
\href{https://journals.aps.org/prd/pdf/10.1103/PhysRevD.108.013002}{Phys. Rev. D \textbf{108}, 013002 (2023)}.

\bibitem{Collas:2018jfx}
P.~Collas and D.~Klein,
\newblock {\em The Dirac Equation in Curved Spacetime: A Guide for Calculations} (Springer, Cham, Switzerland, 2019).

\bibitem{Bar:2009zzb}
C.~B\"ar and K.~Fredenhagen,
Quantum field theory on curved spacetimes: Concepts and Mathematical Foundations,
\href{https://link.springer.com/book/10.1007/978-3-642-02780-2}{Lect. Notes Phys. \textbf{786}, 1 (2009)}.

\bibitem{Saulson:2010zz}
P.~R.~Saulson,
Josh Goldberg and the physical reality of gravitational waves,
\href{https://link.springer.com/article/10.1007/s10714-011-1237-z}{Gen. Rel. Grav. \textbf{43}, 3289 (2011)}.

\bibitem{LIGOScientific:2016aoc}
B.~P.~Abbott \textit{et al.} [LIGO Scientific and Virgo],
Observation of Gravitational Waves from a Binary Black Hole Merger,
\href{https://journals.aps.org/prl/pdf/10.1103/PhysRevLett.116.061102}{Phys. Rev. Lett. \textbf{116}, 061102 (2016)}.

\bibitem{Maggiore:2008}
M. Maggiore,
\newblock {\em Gravitational Waves: Volume 1: Theory and
Experiments} (Oxford University Press, New York, Switzerland, 2008), Vol. 1.

\bibitem{Einstein:1916vd}
A.~Einstein,
The foundation of the general theory of relativity,
\href{https://onlinelibrary.wiley.com/doi/10.1002/andp.19163540702}{Annalen Phys. \textbf{49}, 769 (1916)}.

\bibitem{Sobol:2020frh}
O.~O.~Sobol, E.~V.~Gorbar, A.~I.~Momot, and S.~I.~Vilchinskii,
Schwinger production of scalar particles during and after inflation from the first principles,
\href{https://journals.aps.org/prd/pdf/10.1103/PhysRevD.102.023506}{Phys. Rev. D \textbf{102}, 023506 (2020)}.

\bibitem{Berry:1984jv}
M.~V.~Berry,
Quantal phase factors accompanying adiabatic changes,
\href{https://royalsocietypublishing.org/doi/10.1098/rspa.1984.0023}{Proc. Roy. Soc. Lond. A \textbf{392}, 45 (1984)}.

\bibitem{Penzias:1965wn}
A.~A.~Penzias and R.~W.~Wilson,
A Measurement of excess antenna temperature at 4080-Mc/s,
\href{https://articles.adsabs.harvard.edu/pdf/1965ApJ...142..419P}{Astrophys. J. \textbf{142}, 419 (1965)}.

\bibitem{BICEP2:2014owc}
P.~A.~R.~Ade \textit{et al.} [BICEP2],
Detection of $B$-Mode Polarization at Degree Angular Scales by BICEP2,

\bibitem{Ida:2002ez}
D.~Ida, K.~y.~Oda, and S.~C.~Park,
Rotating black holes at future colliders: Greybody factors for brane fields,
\href{https://journals.aps.org/prd/pdf/10.1103/PhysRevD.67.064025}{Phys. Rev. D \textbf{67}, 064025 (2003) [erratum: Phys. Rev. D \textbf{69}, 049901 (2004)]}.

\bibitem{Zhang:2024xod}
X.~L.~Zhang, Y.~F.~Huang, and Z.~C.~Zou,
Recent progresses in strange quark stars,
\href{https://arxiv.org/pdf/2404.00363}{[arXiv:2404.00363 [quant-ph]]}.

\bibitem{Huang:1997}
Y.~F.~Huang and T.~Lu,
\href{https://articles.adsabs.harvard.edu/pdf/1997A%26A...325..189H}{A\&A, \textbf{325}, 189 (1997)}.

\bibitem{Madsen:1998qb}
J.~Madsen,
How to identify a strange star,
\href{https://journals.aps.org/prl/pdf/10.1103/PhysRevLett.81.3311}{Phys. Rev. Lett. \textbf{81}, 3311 (1998)}.

\bibitem{Glendenning:1989cf}
N.~K.~Glendenning,
Fast Pulsar in {SN1987A}: Candidate for Strange Quark Matter,
\href{https://journals.aps.org/prl/pdf/10.1103/PhysRevLett.63.2629}{Phys. Rev. Lett. \textbf{63}, 2629 (1989)}.

\bibitem{Dai:2015jwa}
Z.~G.~Dai, S.~Q.~Wang, J.~S.~Wang, L.~J.~Wang, and Y.~W.~Yu,
The Most Luminous Supernova ASASSN-15lh: Signature of a Newborn Rapidly-Rotating Strange Quark Star,
\href{https://iopscience.iop.org/article/10.3847/0004-637X/817/2/132/pdf}{Astrophys. J. \textbf{817}, 132 (2016)}.

\bibitem{Ford:1986sy}
L.~H.~Ford,
Gravitational Particle Creation and Inflation,
\href{https://journals.aps.org/prd/pdf/10.1103/PhysRevD.35.2955}{Phys. Rev. D \textbf{35}, 2955 (1987)}.

\bibitem{Wu:2019kyb}
Q.~Wu, W.~Zhu, and L.~Feng,
Testing the Wave-Particle Duality of Gravitational Wave Using the Spin-Orbital-Hall Effect of Structured Light,
\href{https://www.mdpi.com/2218-1997/8/10/535}{Universe \textbf{8}, 535 (2022)}.

\bibitem{Li:2019rex}
Z.~L.~Li, B.~S.~Xie, and Y.~J.~Li,
Boson pair production in arbitrarily polarized electric fields,
\href{https://journals.aps.org/prd/pdf/10.1103/PhysRevD.100.076018}{Phys. Rev. D \textbf{100}, 076018 (2019)}.

\end{thebibliography}
\end{document}